\documentclass[review]{elsarticle}
\usepackage{hyperref}
\usepackage{placeins}
\usepackage{subfigure}
\DeclareGraphicsExtensions{.pdf,.jpeg,.png,.jpg,.eps}

\journal{Journal of Physica B}

\begin{document}

\begin{frontmatter}

%\title{Elsevier \LaTeX\ template\tnoteref{mytitlenote}}
\title{Simulation Tools for Detector and Instrument Design}

\author[ess]{Kalliopi Kanaki\corref{mycorrespondingauthor}}
\cortext[mycorrespondingauthor]{Corresponding author}
\ead{Kalliopi.Kanaki@esss.se}

\author[ess]{Thomas Kittelmann}
\author[ess,dtu]{Xiao Xiao Cai}
\author[dtu,ess]{Esben Klinkby}
\author[dtu]{Erik B. Knudsen}
\author[dtu,ess]{Peter Willendrup}
\author[ess,mid]{Richard Hall--Wilton}

\address[ess]{European Spallation Source ERIC, 22-100, Lund, Sweden}
\address[dtu]{Technical University of Denmark, DTU, 2800 Kgs. Lyngby, Denmark}
\address[mid]{Mid-Sweden University, SE-851 70 Sundsvall, Sweden}

%% \author{Elsevier\fnref{myfootnote}}
%% \address{Radarweg 29, Amsterdam}
%% \fntext[myfootnote]{Since 1880.}

%% %% or include affiliations in footnotes:
%% \author[mymainaddress,mysecondaryaddress]{Elsevier Inc}
%% \ead[url]{www.elsevier.com}

%% \author[mysecondaryaddress]{Global Customer Service\corref{mycorrespondingauthor}}
%% \cortext[mycorrespondingauthor]{Corresponding author}
%% \ead{support@elsevier.com}

%% \address[mymainaddress]{1600 John F Kennedy Boulevard, Philadelphia}
%% \address[mysecondaryaddress]{360 Park Avenue South, New York}

\begin{abstract}
The high performance requirements at the European Spallation Source
have been driving the technological advances on the neutron detector
front. Now more than ever is it important to optimize the design of
detectors and instruments, to fully exploit the ESS source
brilliance. Most of the simulation tools the neutron scattering
community has at their disposal target the instrument optimization
until the sample position, with little focus on detectors. The ESS
Detector Group has extended the capabilities of existing detector
simulation tools to bridge this gap. An extensive software framework
has been developed, enabling efficient and collaborative developments of required simulations and analyses --
based on the use of the \texttt{Geant4} Monte Carlo toolkit, but with extended
physics capabilities where relevant (like for Bragg diffraction of
thermal neutrons in crystals). Furthermore, the \texttt{MCPL} (Monte Carlo Particle Lists) particle data exchange file
format, currently supported for the primary Monte Carlo tools of the
community (\texttt{McStas}, \texttt{Geant4} and \texttt{MCNP}),
facilitates the integration of detector simulations with existing
simulations of instruments using these software packages. These means offer a powerful set of tools to
tailor the detector and instrument design to the instrument application.
\end{abstract}

\begin{keyword}
%\texttt{elsarticle.cls}\sep \LaTeX\sep Elsevier \sep template
%\MSC[2010] 00-01\sep  99-00
crystals, file formats, Monte Carlo simulations, neutron
detector, neutron scattering
\end{keyword}

\end{frontmatter}

%\linenumbers

\section{Introduction}

The neutron scattering community has been investing a large effort in
developing new detector technologies that can tackle the needs of the
upcoming European Spallation Source (ESS)~\cite{esstdr}. The detector designs
are diverse and target different sets of requirements~\cite{oliver,brightnessurl}, e.g.\,high spatial
resolution~\cite{gdgem, multiblade}, high rate
capability~\cite{multiblade,bandgem}, large area
coverage~\cite{multigrid} or combinations thereof. Additionally, as
instrument performance is typically defined by the signal to
background ratio~\cite{natasha}, the ability
to predict and improve instrument backgrounds will enhance the
capability of future instruments. To facilitate and
accelerate the design process, the use of \texttt{Geant4}~\cite{g1,g2,g3} has
been adopted. It is a powerful Monte Carlo simulation toolkit for
the description of particle passage through matter, used by several scientific
communities since decades.

In recent years the capabilities of \texttt{Geant4}
have been extended to include an increasing number of neutron-related
phenomena at lower energies. The ESS Detector Group continues this effort by creating easily integrable tools,
modeling Bragg and inelastic processes in crystalline materials. This way, it is becoming
possible to expand the relevant
functionality of \texttt{Geant4} and other software packages utilized by the
neutron scattering community. A selection of these tools is presented
in this article with focus on their functionality and not their
technical implementation. The intention of these tool-sets is two-fold:
firstly to enhance the capability and accuracy of the simulations;
secondly to lower the entry barrier to utilization of the simulation
codes and ensure their correct usage with modern code management and
validation of the code and standard results.

\section{The \texttt{NCrystal} Project}

\begin{figure}[!h]
  \centering
  %% \subfigure[]{\includegraphics[width=0.13\columnwidth]{MGmodel}
  %%   \label{}
  %% }\hspace{0.1mm}
  \subfigure[]{\includegraphics[width=0.48\columnwidth]{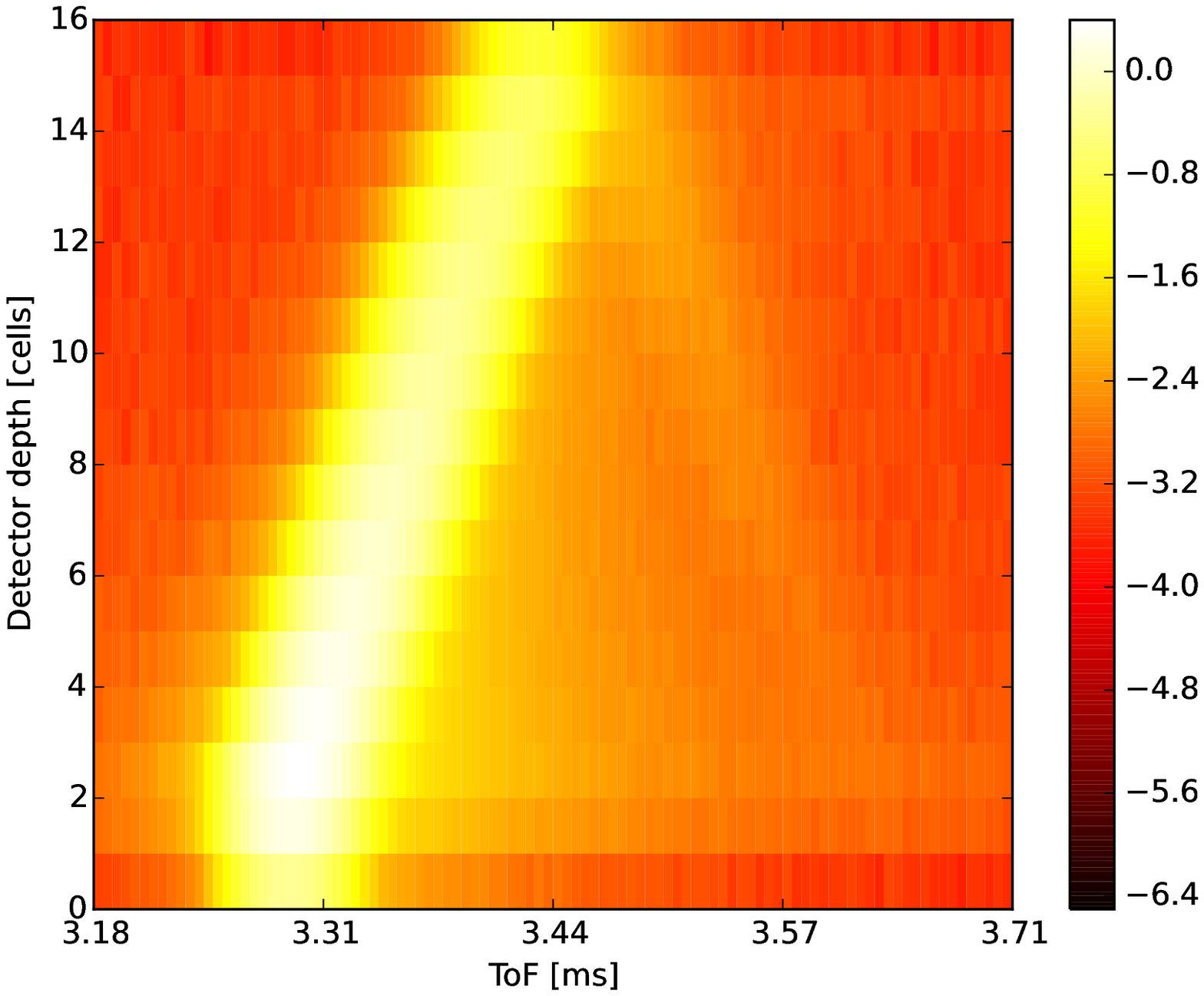}
    \label{}
  }\hspace{0.05mm}
  \subfigure[]{ \includegraphics[width=0.48\columnwidth]{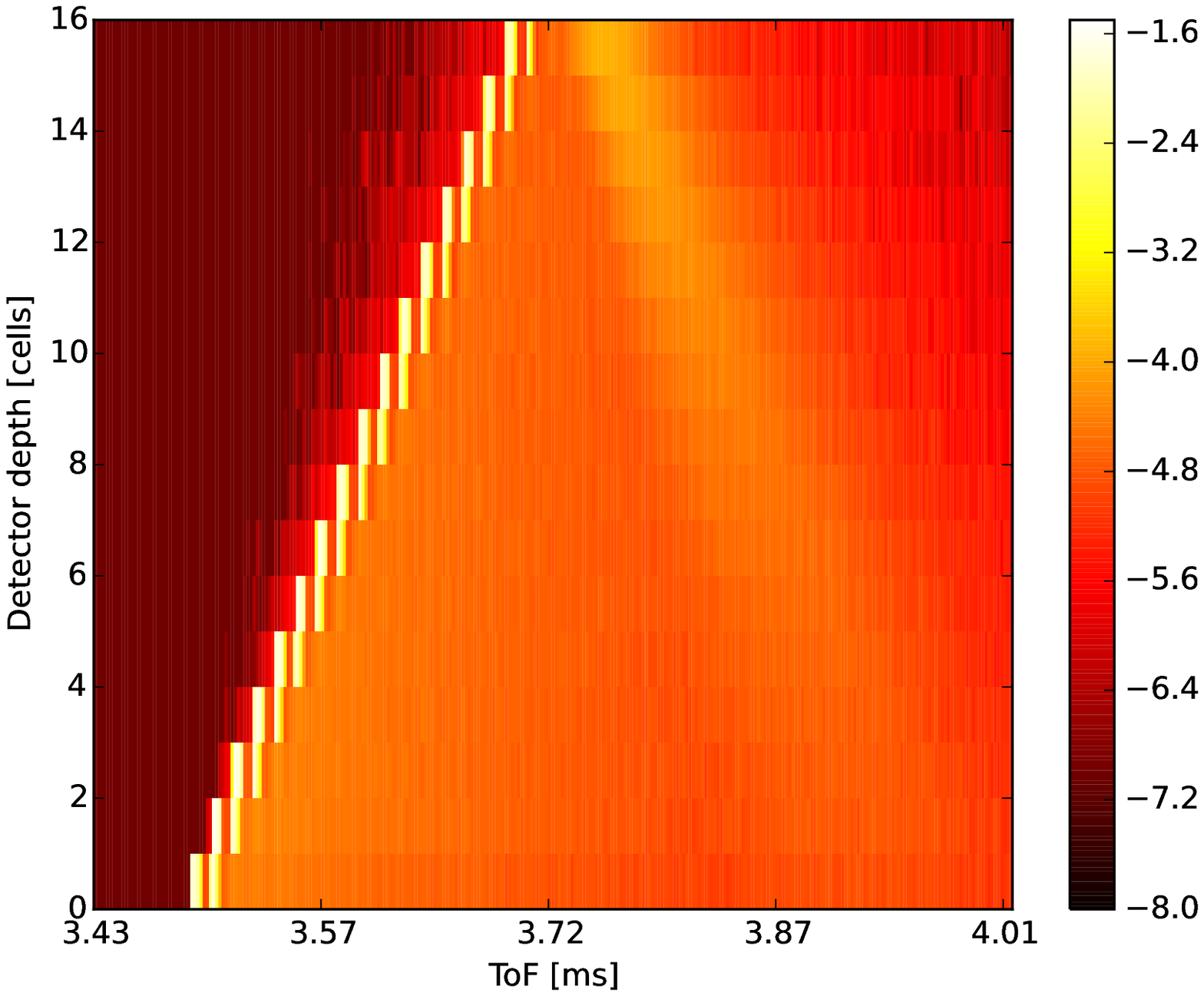}
    \label{}
  } \caption{\label{BS_MG} Back-scattered neutrons at the rear end of
    the Multi-Grid detector module (a) (from~\cite{mgin6}) can be
    reproduced with \texttt{Geant4} (b) (from~\cite{dian_ieee16}), provided the
    coherent scattering for the crystalline Al detector frame is
    enabled via \texttt{NXSG4}.}
\end{figure}

A large fraction of the instrument components consists of
crystalline materials, which makes it crucial for simulations to correctly model interactions
of thermal neutrons in such materials. One
of the first \texttt{Geant4}-extensions written for this purpose is
\texttt{NXSG4}~\cite{nxsg4, nxsg4url}. It provides \texttt{Geant4} with a description of Bragg scattering on a selection of powder materials. It is
extensively used to study the impact of the detector material budget
on the simulated detector signal. Fig.~\ref{BS_MG} demonstrates how
such an effect can be reproduced by enabling the crystal properties of
Aluminium in a \texttt{Geant4} simulation of the Multi-Grid
detector~\cite{mgin6,dian_ieee16} via \texttt{NXSG4}.

With \texttt{NXSG4} being a precursor, further advances on this front aim
to expand the material functionalities, including the treatment of
scattering on single-crystals, compound materials and non-Bragg
processes. This effort is combined in the \texttt{NCrystal} project, which is scheduled
for the first public release in August 2017 at \cite{ncrystalgithub}, and which will
come with appropriate extensions for integration with both
\texttt{McStas}~\cite{mcstas} and \texttt{Geant4}.

\begin{figure}[!h]
\centering
\includegraphics[width=0.65\columnwidth]{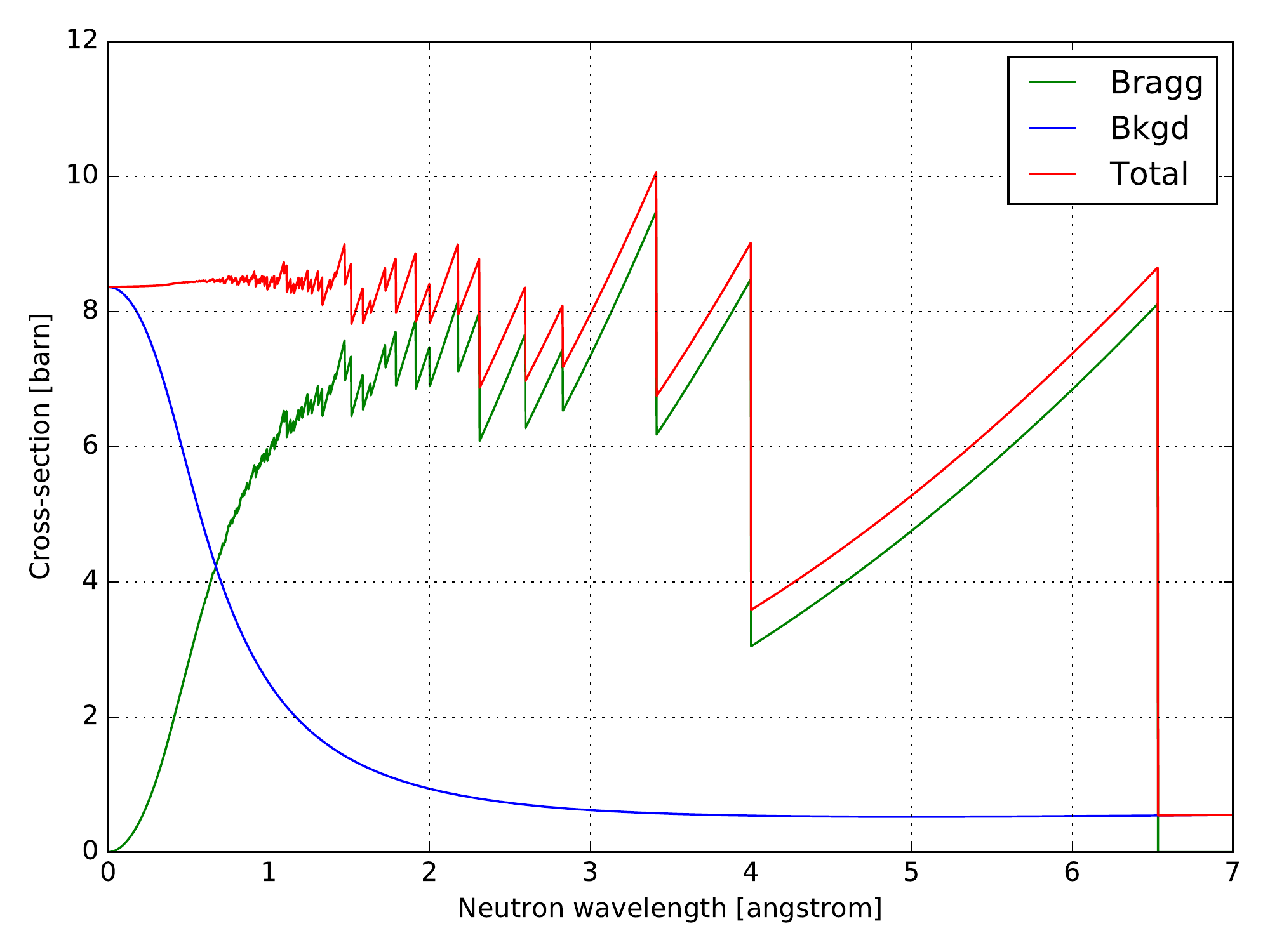}
\caption{\footnotesize Total neutron cross sections for Germanium powder as a
  function of neutron wavelength, overlaid with the Bragg and
  background contributions, provided by \texttt{NCrystal}.}
\label{Ge_xsect}
\end{figure}

\begin{figure}[!h]
\centering
\includegraphics[width=0.65\columnwidth]{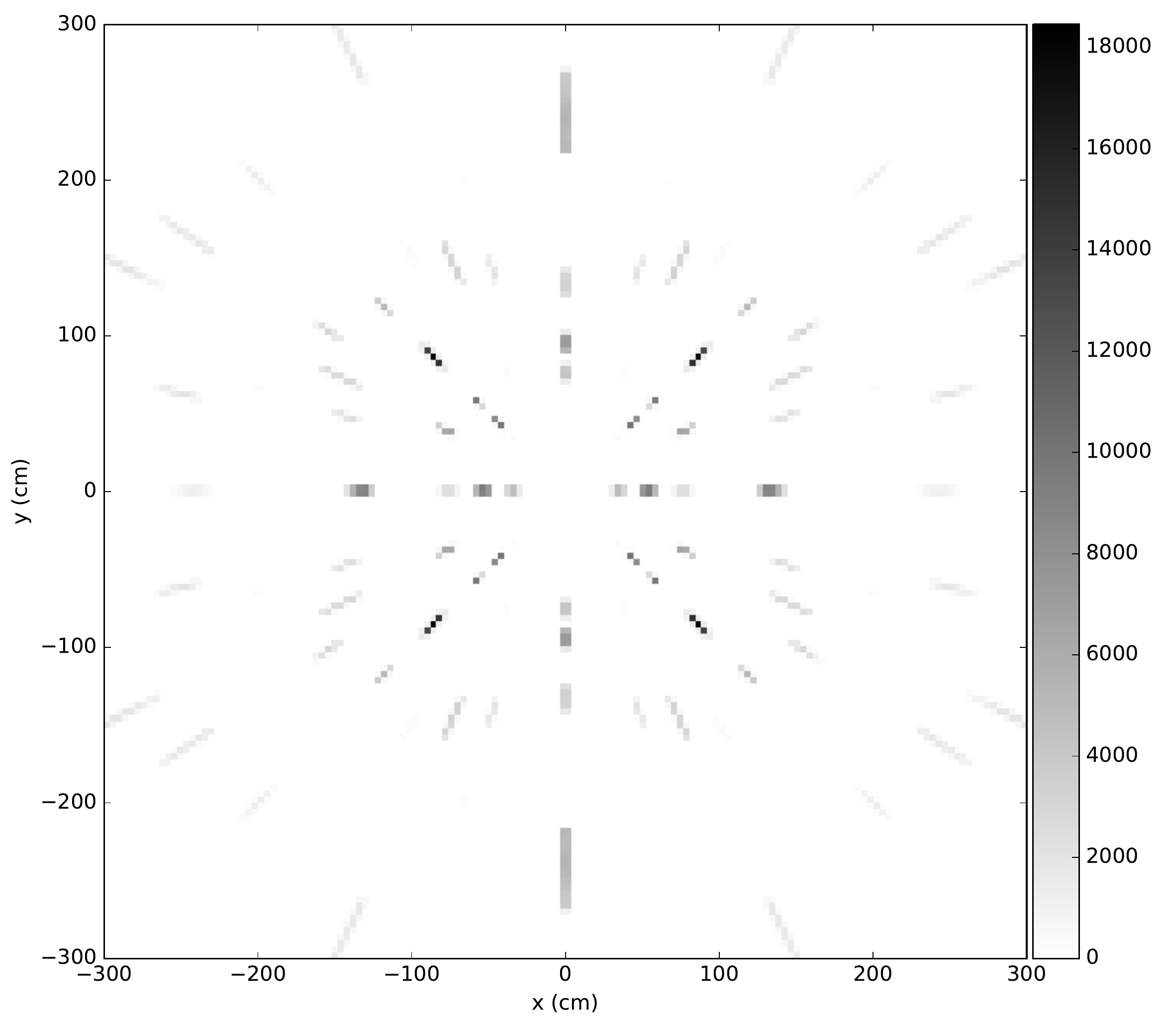}
\caption{\footnotesize Bragg diffraction pattern from a single
  buckminsterfullerene (C$_{60}$) structure, as simulated by \texttt{NCrystal}. The
  pattern is derived with a fixed-orientation white neutron beam
  hitting the crystal. Only Bragg scattering is enabled, while
  multiple scattering is ignored. The generated Bragg pattern shows
  the crystal reciprocal lattice, which is directly comparable with
  those obtained in neutron and X-ray measurements in a similar geometry setup.}
\label{C60laue}
\end{figure}

\begin{figure}[h]
\centering
\includegraphics[width=0.65\columnwidth]{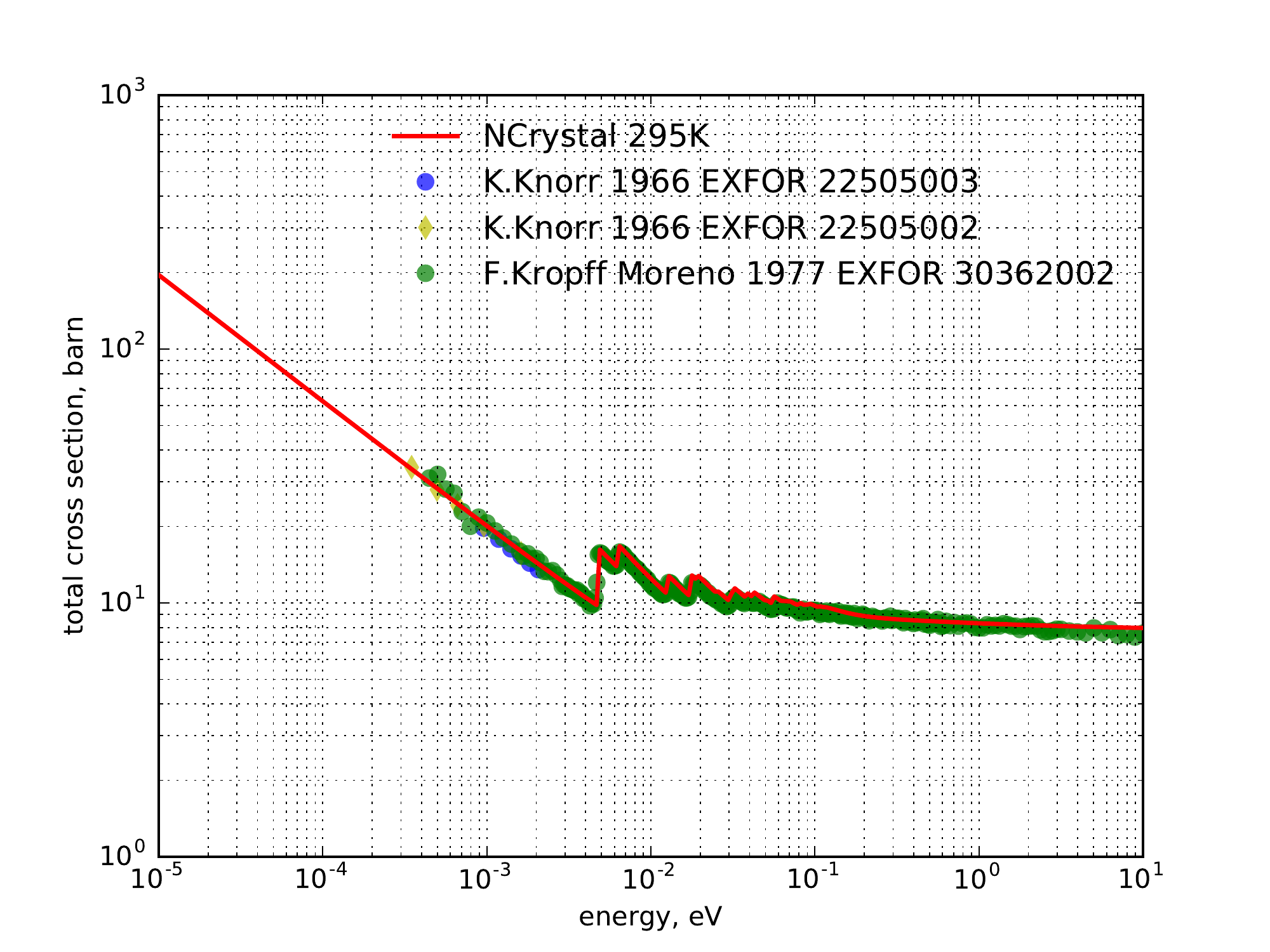}
\caption{\footnotesize Cu total cross section vs.\,neutron
  energy as predicted by \texttt{NCrystal} (red). The experimentally
  determined points are taken from the IAEA EXFOR
  database~\cite{exfor1,exfor2}.}
\label{Cu}
\end{figure}

In the current state of the code, supported are Bragg scattering on
powders and single crystals, as well as an improved description of
inelastic and incoherent processes. A demonstration of its potential
is presented in Figs.~\ref{Ge_xsect} and~\ref{C60laue} for a Germanium powder and a
single mosaic C$_{60}$ crystal respectively.
%An improvement on the latter is prepared for \texttt{NCrystal} 2.0,
%based on sampling of scattering kernels.
The code is benchmarked against experimental data for a popular list of materials, both single element (e.g.\,Al,
Cu, Ge, Si, Be, V, Pb, C) and compound (e.g.\,CuO, MgO, Al$_2$O$_3$,
SiO$_2$). Such a validation example is shown in Fig.~\ref{Cu} for the modeling of neutron interactions in a Cu powder.

%% A more complex validation example is that of the PUS diffractometer at
%% the IFE reactor in Norway. The instrument has been routinely calibrated with
%% an $\alpha$-Al$_2$O$_3$ sample and a diffraction pattern is recorded (see Fig.~\ref{pus}). \texttt{NCrystal} is used to model the collimators
%% before and after the sample, the sample itself, the Ge-511 monochromator,
%% as well as the shielding between sample and
%% monochromator. The predicted \texttt{NCrystal} diffraction pattern coincides with the measured
%% data points in terms of intensities, widths and peak positions. The discrepancies are understood and can be attributed to
%% the detector resolution not being included in the model. Also, the
%% disagreement at small angles is related to the description of the
%% inelastic background and is anticipated to improve with a future
%% \texttt{NCrystal} release, which will focus on improving the modeling
%% of the inelastic contribution.
%% \begin{figure}[!h]
%% \centering
%% \includegraphics[scale=0.3]{pus}
%% \caption{\footnotesize Diffraction pattern recorded at the PUS@IFE
%%   diffractometer (black points). In red appears the \texttt{NCrystal} prediction,
%%   demonstrating the fine agreement with the measured spectrum.}
%% \label{pus}
%% \end{figure}

\texttt{NCrystal} offers accurate yet efficient descriptions of neutron-crystal
interactions, arguably unprecedented in a general-purpose, open source library,
and is expected to play a crucial role in advancing the evaluation of simulated instrument
and detector performance.

\FloatBarrier

\section{The \texttt{MCPL} File Format}

Motivated by the need for simulation packages to efficiently exchange
particle data, the ESS Detector Group developed the Monte Carlo
Particle List (\texttt{MCPL}) file format~\cite{mcplpaper}. It is a
well-defined, binary format containing full particle
state information. Interfaces
are available for \texttt{Geant4}, \texttt{McStas} 2.4.1, \texttt{McXtrace} 1.4~\cite{mcxtrace}, \texttt{MCNP5}~\cite{mcnp}, \texttt{MCNP6} and \texttt{MCNPX}, so
that the communication between these packages is facilitated
through a single standardized format. \texttt{MCPL} comes with C/C++/Python
bindings that allow easy integration with software packages. It
also constitutes a convenient way for storage of particle state information
even for users of just a single software application. The stand-alone code can be downloaded from its GitHub
repository~\cite{mcplgithub}, is open source and released under very liberal
license conditions. The power of the \texttt{MCPL} format lies in its
flexibility and efficient implementation but
also in the utilities it ships with. Easy histogram visualization,
file merging, filtering and modification are available via simple
terminal commands~\cite{pythonmcpl}. The \texttt{MCPL} format allows the most appropriate or familiar simulation code to be used by the user for the application foreseen.

\begin{figure}[!th]
\centering
\includegraphics[width=0.65\columnwidth]{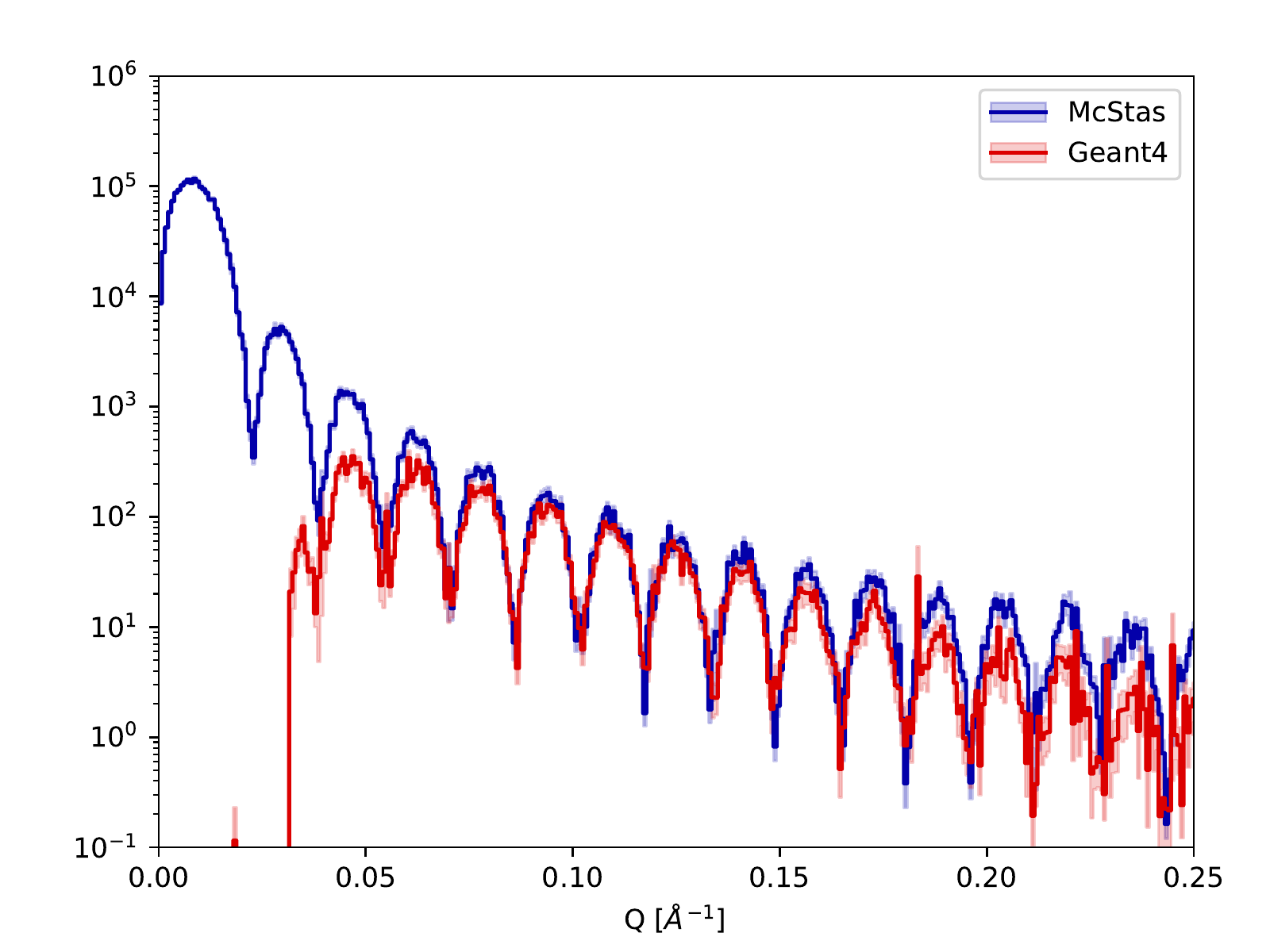}
\caption{\footnotesize Raw Q distribution for a subset of the
  detectors of the instrument (from~\cite{mcplpaper}). The
    \texttt{\texttt{McStas}} post-sample output appears in blue, while the distribution calculated
    from the simulated measurements in \texttt{Geant4} appears in red.}
\label{lokiq}
\end{figure}

As a show case for the \texttt{MCPL} use, the interfacing of a \texttt{McStas} instrument simulation to a
\texttt{Geant4} detector simulation is presented here. As a part of
the detector and instrument optimization processes, it is beneficial
to use customized input for the detector simulation, e.g.\,a realistic
distribution of neutrons scattered from a typical sample for the
particular application. Such a scenario is shown in
Fig.~\ref{lokiq}. Neutrons are simulated in a Small Angle Neutron Scattering instrument in \texttt{McStas} until
after their scattering on 200~\AA\ radius spheres.
These neutrons are then saved with all
their properties in the \texttt{MCPL} format and used as input to a \texttt{Geant4}
simulation, which contains a detailed detector model. This way, the
user is able to look at the interesting scientific quantities both at
the sample and after the detection or any other stage of the
simulation. In this particular example, Fig.~\ref{lokiq} depicts the Q intensity that
contains instrument and sample effects in blue, while in red appears the
result of the \texttt{Geant4} detector simulation, which additionally convolves
the detector effects. 

\FloatBarrier

\section{The ESS Simulation Framework}

It is customary for large scientific communities to create software
frameworks, which provide  support and utilities to all community
members. These should incorporate modern software coding tools such as
a versioned repository, documentation, issue tracking and
validation~\cite{githuburl,confluence,jira,bitbucket}. Such a simulation framework~\cite{dgcode} has been
developed by the ESS Detector Group and is used at the moment within
ESS and by in-kind collaborators in Europe. It facilitates a quick
setup of simulation and analysis of new projects, reducing thus the
overhead required to start a new detector simulation. Users can
benefit from centralized support and utilities that are seamlessly
integrated.
\begin{figure}[!h]
\centering
\includegraphics[width=0.6\columnwidth]{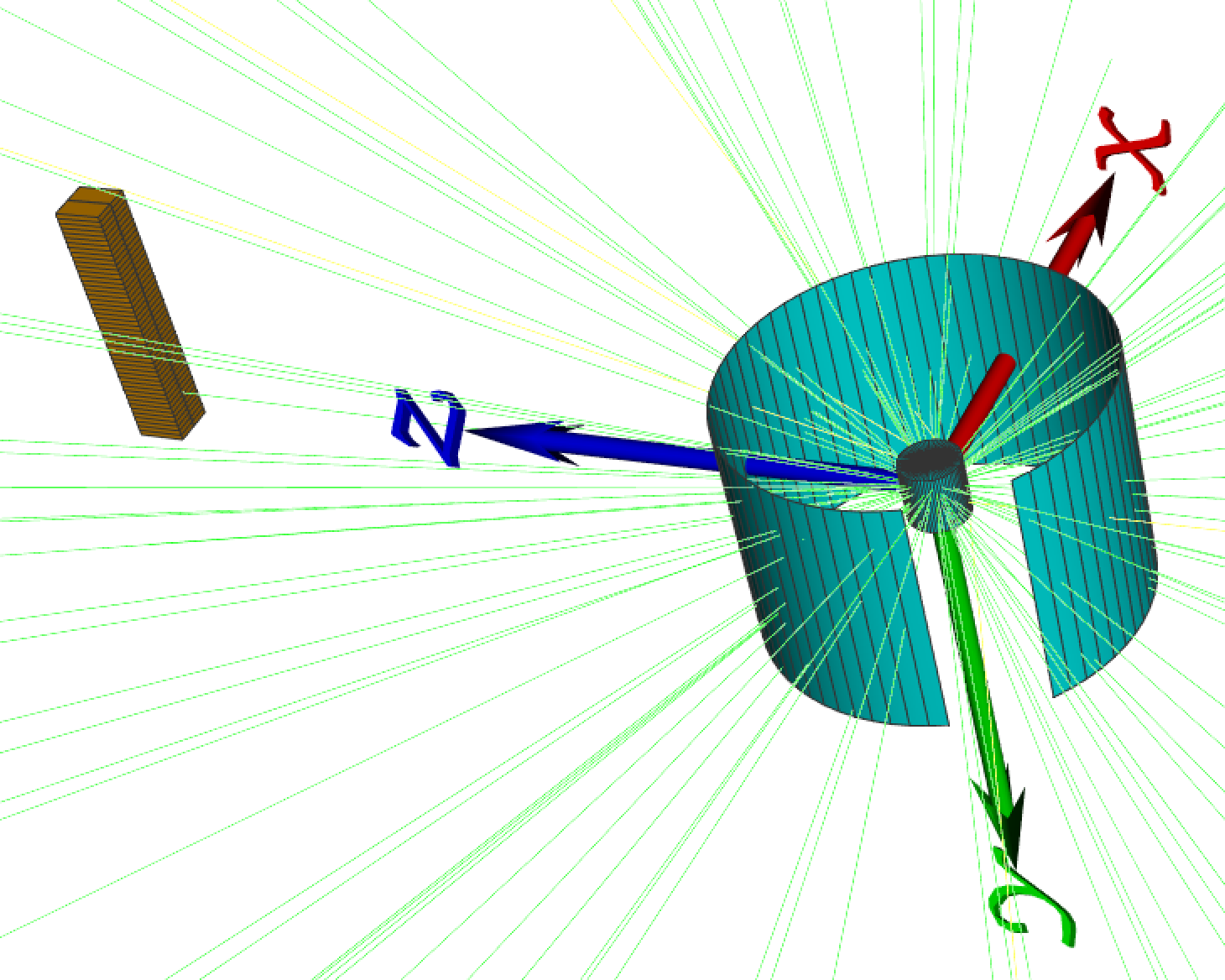}
\caption{\footnotesize Geometry and particle gun visualization with
  the ESS simulation framework viewer. The neutrons are generated at
  the centre of the coordinate system and are emitted isotropically
  towards the Multi-Grid detector modules (in orange). The turquoise
  volumes represent Aluminium components of the sample environment and
  the entrance window.}
\label{viewer}
\end{figure}
To mention just a few of the latter, the framework ships with a
user-friendly, dynamic build system, a flexible
Python interface for combining geometries and particle generators, a
dedicated 3D viewer based on \texttt{OpenSceneGraph}~\cite{osg} that
allows visualization and quick geometry debugging (see Fig.~\ref{viewer}), a customized binary file format (\texttt{GRIFF})~\cite{dgcode} for
efficient storage of simulation output and accompanying meta-data, and
finally NumPy-compatible histogram classes to assist the user with the
analysis of the results. In addition to these utilities, all the aforementioned
libraries (\texttt{NXSG4}, \texttt{NCrystal}, \texttt{MCPL}) are already incorporated and
ready to use.

\section{Outlook}

The instrument and detector requirements at ESS place high demands on the
respective simulation tools to tackle the ambitious design
challenges. With the advances outlined in this article, it is now
possible to accurately model an ever larger fraction of instrument
and detector components. Focus has been given to the inclusion of relevant physics processes, as well as
the communication between the simulation tools of the community,
facilitating the tailoring of the detector designs to the scientific
application. This allows the most appropriate simulation program to be
used for the particular task in-hand.

\section*{Acknowledgments}

This work was supported in part by the EU Horizon 2020 framework,
BrightnESS project 676548.

\end{document}